\documentclass[aip,apl,reprint]{revtex4-1}

\usepackage{graphicx}

\begin{document}

\title{Virtual Scanning Tunneling Microscopy:\\ A Local Spectroscopic Probe of 2D Electron Systems}

\author{A. Sciambi}
\affiliation{Department of Applied Physics, Stanford University, Stanford CA 94305-4045 USA}
\affiliation{Stanford Institute for Materials and Energy Sciences, SLAC National Accelerator Laboratory, 2575 Sand Hill Road, Menlo Park, CA 94025 USA}
\author{M. Pelliccione}
\affiliation{Department of Applied Physics, Stanford University, Stanford CA 94305-4045 USA}
\affiliation{Stanford Institute for Materials and Energy Sciences, SLAC National Accelerator Laboratory, 2575 Sand Hill Road, Menlo Park, CA 94025 USA}
\author{S. R. Bank}
\affiliation{Materials Department, University of California Santa Barbara, Santa Barbara CA 93106 USA}
\affiliation{Electrical and Computer Engineering Department, University of Texas at Austin, Austin TX 78758 USA}
\author{A. C. Gossard}
\affiliation{Materials Department, University of California Santa Barbara, Santa Barbara CA 93106 USA}
\author{D. Goldhaber-Gordon}
\affiliation{Department of Physics, Stanford University, Stanford CA 94305-4045 USA}
\affiliation{Stanford Institute for Materials and Energy Sciences, SLAC National Accelerator Laboratory, 2575 Sand Hill Road, Menlo Park, CA 94025 USA}

\date{\today}

\begin{abstract}
We propose a novel probe technique capable of performing local low-temperature spectroscopy on a 2D electron system (2DES) in a semiconductor heterostructure. Motivated by predicted spatially-structured electron phases, the probe uses a charged metal tip to induce electrons to tunnel locally, directly below the tip, from a ``probe'' 2DES to a ``subject'' 2DES of interest. We test this concept with large-area (non-scanning) tunneling measurements, and predict a high spatial resolution and spectroscopic capability, with minimal influence on the physics in the subject 2DES.
\end{abstract}

\maketitle

As semiconductor growth techniques advance, two-dimensional electron systems (2DESs) with ultra-low disorder are revealing exotic new physics. For instance, anisotropic transport in high mobility quantum Hall systems suggests a striped mixture of electron phases.\cite{Koulakov,Lilly,Sambandamurthy} Transport in other high quality samples shows metal-insulator transitions\cite{Abrahams,Ilani,Baenninger} whose intermediate states are conjectured by some to be microemulsions of metallic and crystallized electron phases driven by strong electron interactions.\cite{Jamei} Unfortunately, the exact organization of such phases is hard to discern with spatially-averaged transport measurements. Scanned probes have greatly enhanced our understanding of 2DESs\cite{Topinka,Ilani2,Chakraborty,Steele} but have yet to directly map the mixed phases. One hindrance is that the transport evidence for these phases appears only in the lowest-disorder 2DESs, which reside at interfaces at least a hundred nanometers from the surface, behind a large Schottky barrier.

\begin{figure}[hb]
\includegraphics[width=7.5cm]{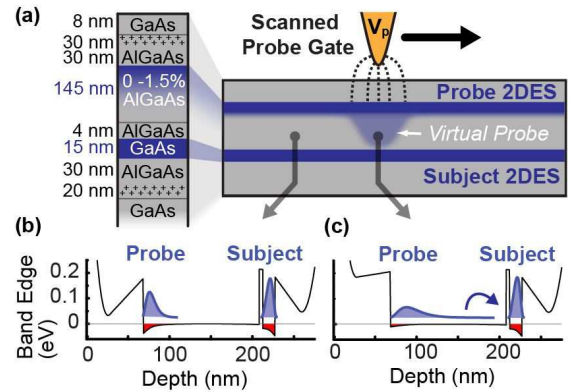} 
\caption{(a) Right: VSTM schematic shows a scanned, charged tip inducing local tunneling from a ``probe'' 2DES into the ``subject'' 2DES. \cite{Eisenstein} Left: Layer composition of heterostructure. (b) Simulated conduction band edge (black), having quantum wells separated by a low and wide barrier, with calculated bound wavefunctions (blue)\cite{Snider}. (c) A negative probe gate voltage (-0.3 V) induces tunneling by pushing the probe layer into the barrier to overlap the subject layer.}
\label{Fig1}
\end{figure}

Scanning tunneling microscopy (STM), the dominant probe of local electronic properties on surfaces,\cite{STM} would be a powerful tool for studying buried semiconductor 2DESs were the tip close enough to allow tunneling. With this in mind, we suggest a novel probe where tunneling comes not from a metal tip but rather from a ``probe'' 2DES grown above the ``subject'' 2DES (Fig 1a). The weakly-coupled 2DESs are separated by a wide but very low potential barrier (Fig. 1b). The barrier is so low that we can induce interlayer tunneling by applying a voltage to a surface gate, modifying the interlayer barrier and thus allowing the wave function of the probe 2DES to penetrate deeper into the barrier. Were this surface gate replaced by a metallic tip scanned above, inter-2DES tunneling would occur preferentially below the tip, forming a ``virtual tip'' that moved with the physical tip. Hence, we call this proposed method Virtual STM (VSTM).

The novelty of this probe is not in its scanning system, which is well established, but in its heterostructure that permits modulation of tunneling between layers. We have recently developed and characterized such a heterostructure, incorporating it into a non-scanning device: the Wavefunction Extension Transistor (WET).\cite{VSTMtrans} The heterostructure (Fig 1a, left) uses a 145 nm-wide AlGaAs barrier with an Al gradient of 0-1.5\% from top to bottom, and is described in more detail in Ref. 13. In this letter, we use a WET as proof of the VSTM principle, implementing fixed, large-area ($\sim 10^5\mu$m$^2$) surface gates in place the scanned VSTM tip. Based on simulation and WET measurements, we predict fully-realized VSTM will have high tunneling spatial resolution, spectroscopic capability, and little probe influence on the subject 2DES. Before discussing VSTM attributes, we briefly address tunnel modulation, the central effect giving rise to the WET and VSTM.

Induced tunneling is a dramatic increase in inter-2DES tunneling conductance brought about by nearly depleting the probe layer with a negatively-biased gate. A negative voltage raises the band edge in the subband of the probe layer, effectively lowering the relative barrier height. This causes the probe wavefunction to extend into the barrier (Fig 1c) and overlap more with the subject layer. This is seen empirically as an increase in the tunneling: for temperatures below 1K, gating the WET can raise tunnel conductance by more than two orders of magnitude (Fig 2a). Because tunnel modulation relies on the extent of the probing wavefunctions and not on the subject layer, it is present across a wide range of subject 2DES densities in the WET. Furthermore, the tunnel modulation is observed to grow roughly exponentially as the probe 2DES is depleted, though reproducible tunnel resonances likely related to quantum interference in the barrier region\cite{THETA} are superimposed on this behavior.

\begin{figure}[bp]
\includegraphics[width=7.5cm]{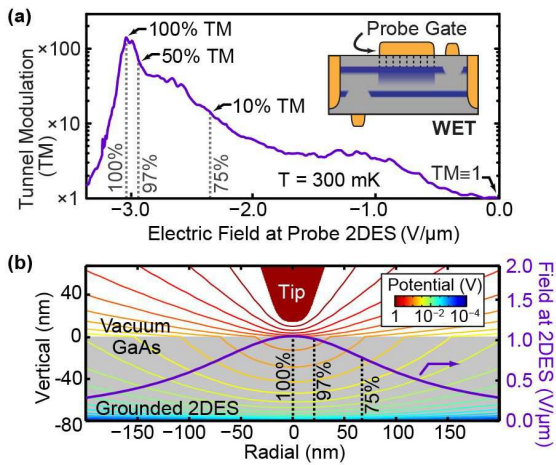} 
\caption{(a) Induced tunneling measured in a WET (inset), as a function of the calculated electric field from the probe gate at the probe layer. Tunnel conductance increases by over two orders of magnitude from no field to the optimal field, and is at less than half its peak value at 97\% of this field. By calculating the potential profile of a (b) tip suspended in vacuum over a grounded 2DES in GaAs, we find where the tip electric field at the 2DES (purple) is similarly reduced to 97\%. There, tunneling should be more than halved, giving us an estimate of VSTM resolution of 40 nm.}
\label{FigB}
\end{figure}

This strong dependence of tunneling on gate voltage should give VSTM a high spatial resolution, similar to the high resolution of STM that arises from sensitivity to tip proximity. In case of VSTM, the falloff of the electric field away from the tip at the probe 2DES should be matched by a sharp decline in tunneling. We can estimate this resolution using large-area tunneling measurements. The negatively-gated WET has its peak value of tunnel conductance halved when the field at the probe layer is only slightly reduced to 97\% of its optimal value (Fig 2a). This reduction can be compared to the modeled field profile of a tip with a 20 nm radius positioned 20 nm above a sample surface, and a further 80 nm above a grounded probe 2DES (Fig 2b). We calculate that the field at the probe 2DES from such a tip falls to 97\% at 20 nm off-center, yielding a tunnel full-width at half-maximum of 40 nm. This serves as a reasonable estimate of the spatial resolution, a value which broadens to 130 nm if tunneling is more conservatively required to fall to one tenth of the maximum.

This analysis relies on a direct relationship between localized field and similarly localized tunneling. For this to be so, the confined tunneling region must have within it a sharp subband edge to give the barrier-penetrating subband wavefunction definite energy and form. It is not obvious, however, how the constituent energy eigenstates of a subband will respond to a sharp potential perturbation in position-space. Our simulations show that for perturbations spatially larger than $\lambda_F$, there is a sharp spectral divide between high-energy states with weight in the perturbed region, and excluded low-energy states. Hence, treating that region as a depleted subband containing only higher-energy states is valid. In a region smaller than $\lambda_F$, we find that all states have weight in the local region and that there is no well-defined subband edge. Using this as a guide, for probe 2DES densities of 2 or 4 $\times$ 10$^{11}$ cm$^{-2}$, the resolution could be as fine as $\lambda_F \approx $ 56 nm or 40 nm, respectively. 

We note we have yet to actually measure tunnel modulation from a small region of 2DES, though for regions larger than $\lambda_F$ the signal size should simply scale with tunneling area. A small local induced tunneling signal should be measurable as long as it is not obscured by weak parallel conduction everywhere else. For a 10 $\mu\mathrm{m}$ $\times$ 10 $\mu\mathrm{m}$ scan area and a 0.1 $\mu\mathrm{m}$ $\times$ 0.1 $\mu\mathrm{m}$ modulated area, the modulated signal at 300 mK should be a 1\% change against a 1 G$\Omega$ background, assuming a tip makes the same 100-fold increase locally that we see for the large-area WET in Fig 2a. Another WET device with a narrower interlayer barrier would yield a 1\% change on a 100 M$\Omega$ background based on large area measurements. For this device, a momentum conservation tunneling resonance is seen in large-area devices but this resonance should not be present for tip-induced tunneling. The small expected changes in tunneling can be observed by oscillating the tip voltages and using standard lock-in techniques.

An important concern beyond resolution and signal strength is that probing might perturb the physics being studied. For VSTM, we want the electric field from the VSTM tip to modify the probe layer without changing the carrier density of the subject 2DES. In bilayers with narrower interlayer barriers than ours, this independent gating of 2DESs is not always easy.\cite{Spielman} We evaluate the influence on the subject 2DES by varying a WET probe gate voltage ($V_p$) and tracking the subject density as measured by Shubnikov-de Haas (SdH) oscillations in longitudinal conductance (Fig 3a). For each $V_p$, we Fourier transform the oscillations to find a period in inverse perpendicular magnetic field (B), indicating a sheet density $n=(2e/h)[\Delta(1/B)]^{-1}$. We see that the populated probe layer (region 1) screens the probe gate from the subject layer and leaves the density of the latter constant to within $1\%$.

\begin{figure}[tp]
\includegraphics[width=7.5cm]{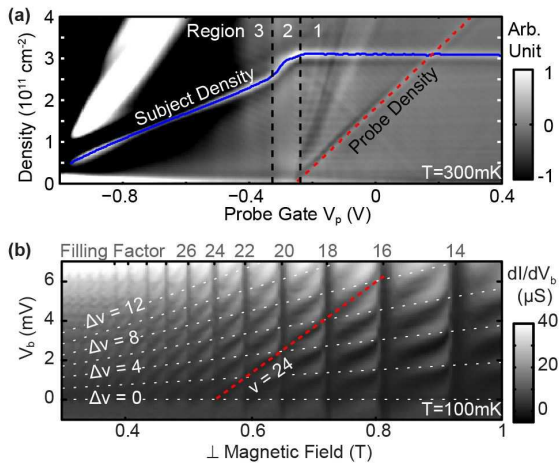} 
\caption{(a) Fourier transform of SdH oscillations reveals the sheet density of each layer as a function of $V_p$. In region 1, the subject layer density is constant to within $1\%$ due to screening by the probe layer. More negative $V_p$ couples to the subject 2DES since the probe 2DES is localized (region 2) or fully depleted (region 3). (b) Spectroscopy of excited/filled Landau levels (filling factor $\Delta\nu=0,4,8,12$) using an interlayer bias ($V_b$). The spin-degenerate Landau level spacings of the density-matched layers (white lines) are linear in field as expected, and a single level ($\nu=24$) is also shown.}
\label{Fig3}
\end{figure}

The SdH oscillations, a result of Landau levels (LLs) appearing in an otherwise flat 2DES density of states (DOS), can also be used to test out-of-equilibrium spectroscopy in the sample. To simplify the measurement, we use the probe gate to match the layer densities and hence level filling. By applying a bias between layers, we can tunnel from filled into excited LLs (Fig 3b). We confirm that the LL spacing is linear in field and we see up to the twelfth excited/filled filling factor, comparable to states probed by capacitive measurements of tunneling into a 2D quantum Hall system from a 3D electrode.\cite{Dial}

The large-area gate characterization of a WET indicates VSTM is in good position to see complex electron phases. To further help, the probe layer can be given a much higher density than the subject layer to smooth the DOS of the former and weaken its interactions. Unlike capacitive measurements of 2DES tunneling, which have enabled impressive spectroscopy of localized and delocalized states,\cite{Steele,Dial} VSTM will need an exit path for the tunneling current to be measured. Thus, it can map the borders of large regions of localized electrons but not the local DOS within. With finite bias, however, electrons can be injected into localized states from which they could then escape into nearby metallic or delocalized states.

We have shown many benefits of VSTM, namely its spectroscopic capability, predicted high spatial resolution, and minimal effect on the subject 2DES. These features are all due to the novel heterostructure design with its especially wide and low barrier. The design gives versatility too, as the VSTM may also be used to map wave functions inside lithographically-defined structures like quantum point contacts and quantum dots. With future heterostructure design iterations, it should be possible to greatly increase tunnel modulation and mobility, though VSTM appears a viable probe technique even with the samples at hand.

We thank C.X. Liu for theoretical discussions, and M.P. Lilly for help with bilayer device fabrication. This work is supported by DOE-BES, DMS\&E at SLAC (DE-AC02-76SF00515), with the original concept developed under the Center for Probing the Nanoscale (NSF NSEC Grant No. 0425897) and a Mel Schwartz Fellowship from the Stanford Physics Department. This work was performed, in part, at the Center for Integrated Nanotechnologies, a DOE-BES user facility at Sandia National Labs (DE-AC04-94AL85000). A.S. acknowledges support from an NSF Fellowship, and M.P. from a Hertz Fellowship, an NSF Fellowship, and a Stanford Graduate Fellowship. D.G.-G. recognizes support from a David and Lucile Packard Fellowship.


\begin{thebibliography}{50}
{
\footnotesize
\bibitem{Koulakov} A. A. Koulakov, M. M. Fogler, and B. I. Shklovskii, Phys. Rev. Lett. \textbf{76}, 499 (1996).
\bibitem{Lilly} M. P. Lilly, K. B. Cooper, J. P. Eisenstein, L. N. Pfeiffer, and K. W. West, Phys. Rev. Lett. \textbf{82}, 394 (1999).
\bibitem{Sambandamurthy}G. Sambandamurthy, R. M. Lewis, Han Zhu, Y. P. Chen, L. W. Engel, D. C. Tsui, L. N. Pfeiffer, and K. W. West, Phys. Rev. Lett. \textbf{100}, 256801 (2008).
\bibitem{Abrahams} A metal-insulator transition review: E. Abrahams, S. V. Kravchenko, and M. P. Sarachik, Rev. Mod. Phys. \textbf{73}, 251 (2001).
\bibitem{Ilani} S. Ilani, A. Yacoby, D. Mahalu, and H. Shtrikman, Science \textbf{292}, 1354 (2001).
\bibitem{Baenninger} M. Baenninger, A. Ghosh, M. Pepper, H. E. Beere, I. Farrer, and D. A. Ritchie, Phys. Rev. Lett. \textbf{100}, 016805 (2008).
\bibitem{Jamei} R. Jamei, S. Kivelson, and B. Spivak, Phys. Rev. Lett. \textbf{94}, 056805 (2005).
\bibitem{Topinka} M. A. Topinka, B. J. LeRoy, R. M. Westervelt, S. E. J. Shaw,
R. Fleischmann, E. J. Heller, K. D. Maranowski, and A. C. Gossard, Nature \textbf{410}, 183 (2001).
\bibitem{Chakraborty} S. Chakraborty, I. J. Maasilta, S. H. Tessmer, and M. R. Melloch, Phys. Rev. B \textbf{69}, 073308 (2004). 
\bibitem{Ilani2} S. Ilani, J. Martin, E. Teitelbaum, J. H. Smet, D. Mahalu,
V. Umansky, and A. Yacoby, Nature \textbf{427}, 328 (2004).
\bibitem{Steele} G. A. Steele, R. C. Ashoori, L. N. Pfeiffer, and K. W. West, Phys. Rev. Lett. \textbf{95}, 136804 (2005).
\bibitem{STM} H.-J. G\"untherodt and R. Wiesendanger (eds.), Scanning
Tunneling Microscopy, Vol I (Springer, 1992).
\bibitem{VSTMtrans} Paper submitted in parallel.
\bibitem{THETA} M. Heiblum and M. V. Fischetti, IBM J. Res. and Develop. \textbf{34}, 530 (1990).
\bibitem{Eisenstein} J. P. Eisenstein, L. N. Pfeiffer, and K. W. West, Appl. Phys. Lett. \textbf{57}, 2324 (1990).
\bibitem{Snider} Simulations used code by G. L. Snider, Univeristy of Notre Dame (http://www.nd.edu/$\sim$gsnider/), and by C.X. Liu, Tsinghua University (unpublished).
\bibitem{Spielman} I. B. Spielman, PhD Thesis, Cal. Tech, pg. 58 (2004).
\bibitem{Dial} O. E. Dial, R. C. Ashoori, L. N. Pfeiffer, and K. W. West, Nature \textbf{448}, 176 (2007).
}
\end{thebibliography}
\end{document}